\documentclass[12pt]{iopart}

\newcommand{\ket}[1]{|#1\rangle}

\usepackage{graphics}
\usepackage{hyperref}
 
\begin{document}

\title{Quantum interference with photon pairs using two micro-structured fibres}

\author{J. Fulconis$^1$, O. Alibart$^1$, W. J. Wadsworth$^2$, and J. G. Rarity$^1$}

\address{$^1$Centre for Communications Research, Department of Electrical and Electronic Engineering\\
University of Bristol, Merchant Venturers Building, Woodland Road, Bristol, BS8 1UB, UK}
\address{$^2$Photonics \& Photonic Materials Group, Department of Physics\\
University of Bath, Claverton Down, Bath, BA2 7AY, UK}
\ead{jeremie.fulconis@bristol.ac.uk}
\begin{abstract}
We demonstrate a quantum interference experiment between two photons coming from non-degenerate pairs created by four-wave mixing in two separated micro-structured fibres. When the two heralded photons are made indistinguishable a $95\%$ visibility is demonstrated.
\end{abstract}

\maketitle

\section{Introduction \label{intro}}
We have developed a high brightness sources of correlated photon pairs using four-wave mixing (FWM) in a photonic crystal fibre (PCF) \cite{fulconisoptexp,alibartnjp}. Bright photon pair sources are essential for future multiphoton experiments \cite{ghz, 4phot, 5phot} and as sources of heralded single photons and entangled states for quantum communications \cite{QKDzbin, QKDwein}. Multiphoton interference is central to the operation of linear optical quantum gates \cite{KLM,cnotrarity,expCNOT} and to the development of optical cluster state computation schemes\cite{Qordi}. All of these experiments are built on interference effects between separate photons. Thus the simplest experimental test of the suitability of a single photon source for multiphoton logic is the demonstration of high visibility interference between separate photons at a beamsplitter. This destructive interference effect was first demonstrated by Hong, Ou and Mandel in 1987 \cite{HOM87, RT88} between photons from degenerate correlated pairs. Later this was demonstrated between heralded  single photons and a weak coherent state \cite{RT96, RTL05} and recently between heralded photons from sources pumped by separate lasers \cite{Zeil06}. The visibility of the interference suppression in all of these experiments is governed by the indistinguishability of the two photons after the beamsplitter. For this the photons must be in a pure state, that is in a single spatial and temporal mode with well defined polarisation. 

In our source the photon pairs are automatically created within a single spatial mode. They are also polarized, narrowband, and extremely bright. However the source is pulsed (photons are $\sim3\,ps$ long) and the natural bandwidth although narrow is not Fourier transform limited. Effectively we start with several coherence times (temporal 'modes') within the pulse length. However suitable band pass filters are chosen to stretch the coherence length to that of the pulse and we are able to show high visibility interference. Previous multi-photon experiments are limited in brightness to less than one four photon event per second mainly due to the limited pump laser power available. Our source is much more efficient and we are able easily to reach hundreds of fourfold coincidences per second. However at these high rates we enter a regime where higher order photon emission is no longer negligible and interference visibility is decreased. 

In the first part of this paper (\sref{sec2}) we briefly describe our photon pair source and how we use it as a heralded single photon source. In the second part (\sref{sec3}), we look at the theoretical expectations of this non-classical interference using FWM compared to other processes such as parametric down conversion. We then focus on the details of an experiment using two spatially separated PCF pair sources and compare our theoretical expectations to the measured results (\sref{sec4}). Finally, we assess our source for quantum logic applications and suggest future improvements required (\sref{disc}).

\section{PCF: from a photon pair source to an heralded single photon source \label{sec2}}
\begin{figure}[h]
\centerline{\scalebox{0.5}{\includegraphics{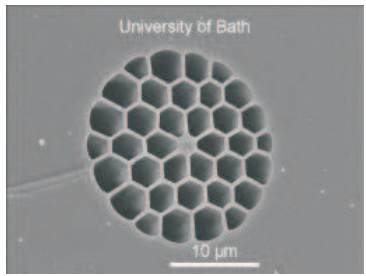}}}
\caption{Electron microscope image of the PCF used with core diameter $d=2\mu m$, $\lambda_0 = 715 nm$ \label {PCF}}
\end{figure}

Quantum interference experiments involving several photons created as two or more pairs require pair photon sources pumped by ultra-short laser pulses where the bandwidth is of order nanometers and also requires a high efficiency of collection \cite{rarityfundpb}. Our source of photon pairs using a PCF appears to be ideal in these terms as it demonstrates all these requirements \cite{fulconisoptexp,alibartnjp}. The main nonlinear process that has to be taken into account is four-wave mixing where two pump photons are converted into a pair of photons following phase matching and conservation of energy:
\begin{eqnarray}  
	&2k_p-k_s-k_i-2\gamma P=&0\\
	&2\omega_p-\omega_s-\omega_i=&0
\end{eqnarray}	
\Fref{PCF} shows the microstructured fibre used in our experiment. It has a zero dispersion point in the near infra-red ($715\,nm$) and is pumped with a picosecond pulsed laser set a few nanometers into the normal dispersion regime at $708\, nm$. 
This generates narrowband polarized photon pairs visible to efficient silicon-based photon counting detectors. The short wavelength sits at $583\, nm$ and the corresponding idler at $900\, nm$ with $6\, nm$ and $11\, nm$ FWHM bandwidths respectively. Due to the guided configuration, our source is exposed to Raman scattering effects. However, the wide separation of the generated pair wavelengths means that the corresponding Raman noise comes from a 6th order process and is thus quite weak. Furthermore, the number of created photon pairs is proportional to the square of the peak intensity while the spontaneous Raman scattering grows roughly linearly (at these low pump powers). Hence, the use of a picosecond laser in our experiment fits the requirements for quantum interference experiments but also allows us to improve the brightness while reducing Raman background noise to negligible levels (less than $10\%$). Consequently, in this configuration and thanks to the high lumped efficiency we could measure up to $2\times10^5$ net coincidences per second with pump power of $1\, mW$ and $20\,cm$ of fibre length. This represents one of the brightest source of photon pairs reported so far. 

\begin{figure}[h]
\centerline{\scalebox{0.6}{\includegraphics{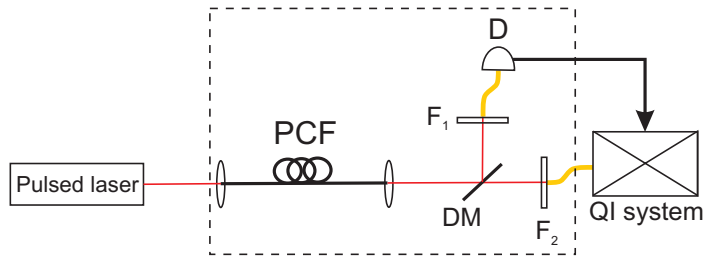}}}
\caption{Heralded single photon source using a PCF fed into a quantum information system.\label {HSP}}
\end{figure}

\begin{figure}[h]
\centerline{\scalebox{0.6}{\includegraphics{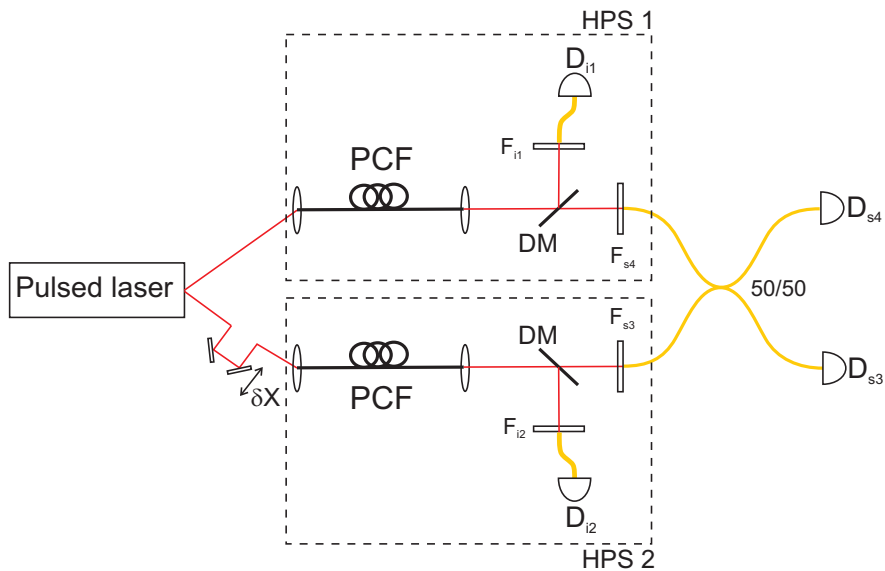}}}
\caption{Simplified setup of the interference experiment using two heralded single photon sources.\label {setupsimple}}
\end{figure}

It has already been demonstrated that photon pair sources can be used to approximate the generation of single photons \cite{pdcmandel, pdcrarity, brightpdc2, spsalibart}. The photons of a pair being emitted simultaneously, the detection of one photon can be used to herald the emission of the second. The setup of an heralded single photon source (HPS) using a PCF is depicted in \fref{HSP}. A pulsed laser is used to pump the fibre. At the output, a dichroic mirror centred at 700 nm divides the pair into two arms, one corresponding to the signal channel and the other to the idler. Each photon of the pair is then launch into a single mode fibre. Once suitably filtered to remove the pump beam, the heralded single photons can then be fed into any quantum communication system (see \fref{HSP}). 
In terms of single photon generation, the coincidence rate reported previously then corresponds to a number of heralded photons detected. These heralded single photon sources play an important role in quantum information applications, however they can be used provided that the emitted photons fit the following requirements. Quantum mechanics predicts indeed that interference effects can be observed when overlapping two independent single photons on a BS as long as they are indistinguishable in spectrum, spatial and temporal modes \cite{rarityfundpb}. This effect is due to the bosonic behavior of photons which says that if two "indistinguishable" photons arrive at the same time on a 50:50 beam-splitter, they will stick together and will always take the same output port. This interference effect can then be observed by looking at the coincidence rate between the two output ports while scanning along the temporal overlap of the photons on the BS. The resulting plot is commonly called the Hong-Ou-Mandel dip, refering to the first experiment of this kind \cite{HOM87}. The visibility of this dip will then give informations on the quality of the indistinguishability, while the width corresponds to the actual coherence length of the interfering photons. To confirm the quality of our source, we thus decided to perform this interference experiment using two separated PCF's and mixed the heralded photons coming from each fibre on a 50:50 BS. This configuration is also a more realistic demonstration of the role that our source could play in multi-photon experiments used in QI systems where the number of single photon sources is getting more and more important. A schematic of the setup is shown on \fref{setupsimple}.
Here, the observation of the two photon interference effect turns into the measurement of the four-fold coincidence rate between pairs created from both fibers. The next section is dedicated to the theoretical study of this non-classical interference experiment.

\section{Quantum interference: theoretical expectations \label{sec3}} 
The origin of the Hong-Ou-Mandel interference effect comes from the indistinguishability and destructive interference between two paths to a coincidindence detection. We can represent a beamsplitter with amplitude transmission and reflection coefficients $t$ and $e^{i\pi/2} r=ir$ ($r,\,t$ real). 
\begin{figure}[h]
\centerline{\scalebox{0.8}{\includegraphics{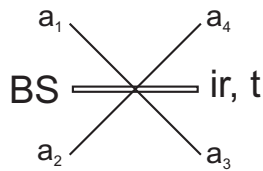}}}
\caption{The input ($a_1$, $a_2$) and output ($a_3$, $a_4$) modes of a beam-splitter (BS).\label {BS}}
\end{figure}
The phase shift on reflection guarantees energy conservation. When we populate the input modes $a$ and $b$ with single photons 
\begin{equation}
|1\rangle_1 |1\rangle_2 \rightarrow (t^2-r^2)|1\rangle_3 |1\rangle_4 +
irt\sqrt{2}(|2\rangle_3|0\rangle_4 +|0\rangle_3 |2\rangle_4).
\end{equation}
When $r=t=1/\sqrt 2$ (the 50/50 beamsplitter) $t^2-r^2=0$ and the
probability of detecting one photon at each output port is zero
\cite{Loudon87} which is clearly non-classical. This effect was first seen
when the coincident photon pairs created in parametric downconversion were
mixed at a beamsplitter \cite{HOM87,RT88}.

The above simple treatment implies that the photons are indistinguishable when viewed from the detectors but contains no spatial or temporal modes explicitly. In the experiment we have a pulsed source of photon pairs with a finite bandwith set by external filters created in a single spatial mode. A more realistic treatment of our interference experiment must include the finite bandwidth and resulting finite temporal overlap of our photons. To do this we follow the treatment first described in \cite {rarityfundpb} where the overlap of heralded photons created in parametric down-conversion (PDC) was studied. The main difference here is that the process of FWM implies two pump-photons are converted into a pair of signal and idler photons. We show in the following, this actually leads to a better interference visibility compared to the experiment using sources based on PDC. Let's start with an effective two-photon wave function at the detectors given by

\begin{equation}
\Psi(\bar{t}_{s},\bar{t}_{i})=\left\langle vac\right|\hat{A}_s(\bar{t}_s) \hat{A}_i(\bar{t}_i)\left|\Psi\right\rangle
\end{equation}
\noindent where $\hat{A}_{s,i}$ are electric field operators for the signal and idler modes ($s,i$). These operators reflect the finite bandwidth and duration of the pairs through,

\begin{equation}
\hat{A}(\bar{t})=\int d\omega f(\omega) e^{i \omega \bar{t}}
\end{equation}

\noindent and reduced time $\bar{t}=t-X/c$ includes the propagation time from a fibre to detectors.\\

\noindent We label the two pump beams $P1,P2$ and detector arms $i1, i2, s3, s4$ as shown in \fref{setupsimple} then before the beamsplitter
\begin{eqnarray}
\Psi_{j,k}(t_{sj},t_{ik})&=&\alpha \int d\omega_{p1}d\omega_{p2}d\omega_{s}d\omega_{i} f_p(\omega_{p1})f_p(\omega_{p2})f_{sj}(\omega_{s})f_{ik}(\omega_{i})\nonumber\\ 
&& \quad \times\delta(\omega_s+\omega_i-\omega_{p1}-\omega_{p2})e^{i(\omega_{sj}t_{sj}+\omega_{ik}t_{ik})}
\end{eqnarray}
with $j=3,4$ and $k=1,2$ and $f_{sj,ik}$ being the function describing the narrowband filters in front of each detector. We identify $\left|\alpha\right|^{2}$ with the probability that a pulse contains a pair (i.e. the mean number of pairs per pulse $\overline n=\left|\alpha\right|^{2}$). We also ignore the vacuum component of the wave-function $\sqrt{1-\left|\alpha\right|^{2}}\ket{vac}$ which completes the normalisation.\\
When only one photon appears in each output, the wave function at detectors $D_{i1}, D_{i2}, D_{s3}, D_{s4}$ is given by 

\begin{equation}
\Psi(i_1,i_2,s_3,s_4)=t^2\Psi_{14}(\bar{t}_{s14},\bar{t}_{i1})\Psi_{23}(\bar{t}_{s23},\bar{t}_{i2})-r^2\Psi_{13}(\bar{t}_{s13},\bar{t}_{i1})\Psi_{24}(\bar{t}_{s24},\bar{t}_{i2}) 
\end{equation}

\noindent The probability of seeing a four-fold coincidence detection is then:
\begin{eqnarray}
P(i_1,i_2,s_3,s_4)&=&\eta_{i1}\eta_{i2}\eta_{s3}\eta_{s4} \int d t_{i1}d t_{i2}d t_{s3}d t_{s4} H(\bar{t}_{i1}-t_0,\Delta t)\nonumber \\
&&\qquad\qquad\qquad H(\bar{t}_{i2}-t_0,\Delta t) H(\bar{t}_{i3}-t_0,\Delta t)\nonumber \\
&&\qquad\qquad\qquad\quad H(\bar{t}_{i4}-t_0,\Delta t) \left|\Psi(i_1,i_2,s_3,s_4)\right|^2 
\end{eqnarray}
\noindent where $\eta_{i1,i2,s3,s4}$ are detector efficiencies and $H(\bar{t}_{l}-t_0,\Delta t)$ $(l=1-4)$ is a normalized detector response function centered on $t_{0}$ that falls to zero when $\bar{t}_{l}-t_0>\Delta t$.\\

The temporal overlap of the two heralded photons on the beam-splitter must be adjusted to within the pump pulse length with fine-tuning $\delta X$ performed on one of the pump beam versus the other one. We thus get
 
\begin{equation}
\bar{t}_{s13}=\bar{t}_{s3}; \qquad \bar{t}_{s23}=\bar{t}_{s3}+\delta X/c; \qquad \bar{t}_{s14}=\bar{t}_{s4}; \qquad \bar{t}_{s24}=\bar{t}_{s4}+\delta X/c
\end{equation}

Following the same calculation steps as in \cite {rarityfundpb} and using identical approximations such as energy-matched Gaussian filter functions where
\begin{eqnarray}
f_p(\omega _p)&=&\frac{1}{N}e^{\frac{(\omega_{p0}-\omega_p)^2}{\sigma_p^2}},\nonumber \\
f_{i1}(\omega_i)&=&\frac{1}{N'}e^{\frac{(\omega_{i0}-\omega_i)^2}{\sigma^2}}=f_{i2}(\omega_i),\nonumber \\
f_{s3}(\omega_s)&=&\frac{1}{N'}e^{\frac{(\omega_{s0}-\omega_s)^2}{\sigma^2}}=f_{s4}(\omega_s),\nonumber \\
2\omega _{p0}&=&\omega _{s0}+\omega _{i0}
\end{eqnarray}
with $\sigma$ the filter bandwidth and $\sigma_p$ the pump bandwidth, we finally get

\begin{equation}\label{prob}
P(i_1,i_2,s_3,s_4)= \overline n^2 \mathcal{N}\left(r^4+t^4-2Vr^2t^2e^{-\frac{\Delta T^2\sigma^2}{2\left(1+\sigma^2/{2\sigma_p^2}\right)}}\right)
\end{equation}
\noindent where $\mathcal{N}$ is a normalising factor, $V$ the interference visibility with: 	 
\begin{equation}\label{vis}
V_{max}=\frac{\sqrt{1+\sigma^2/{\sigma_p^2}}}{1+\sigma^2/{2\sigma_p^2}}.
\end{equation}

This last results reveals that for any given $\sigma, \sigma_p$ the visibility using four-wave mixing is higher than for parametric down-conversion that is given by $V=\frac{\sqrt{1+2\sigma^2/{\sigma_p^2}}}{1+\sigma^2/{\sigma_p^2}}$.
This improved visibility is due to the need for two pump photons to be absorbed in the four-wave mixing process. This reduces the time uncertainty and allows wider band-pass filters to be used to achieve the time-bandwidth limit required.

\section{Quantum interference: experiment and results \label{sec4}}
\begin{figure}[h]
\centerline{\scalebox{0.6}{\includegraphics{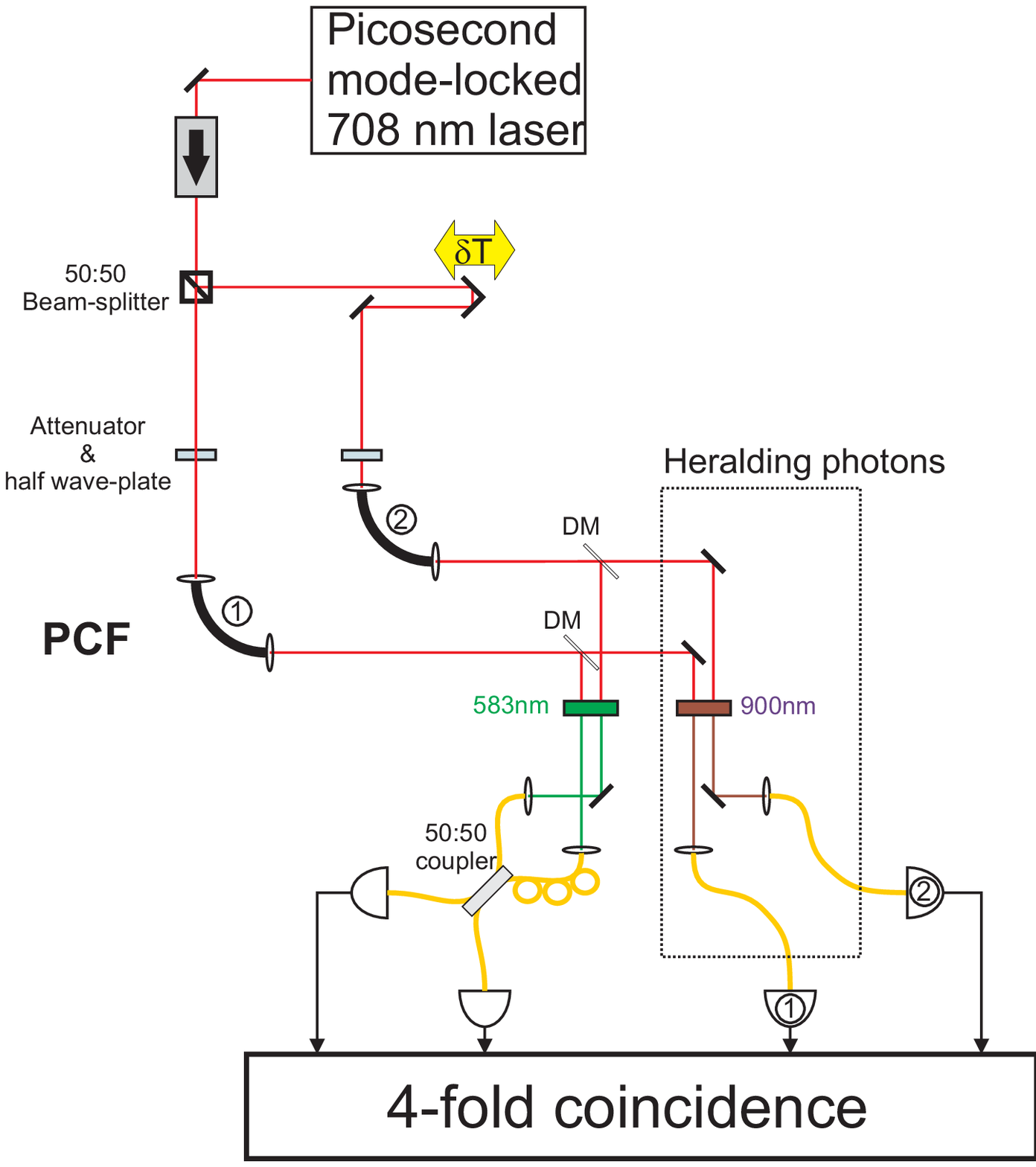}}}
\caption{Setup of the interference experiment.\label {setup}}
\end{figure}

An accurate setup of the experiment is depicted in \fref{setup}. A mode-locked picosecond Ti:Sapphire pump laser (Spectra Physics - Tsunami) set at $708\,nm$, emitting $1.5\, ps$ pulses with a repetition rate of $82\, Mhz$ is sent onto a 50:50 beam-splitter. The output modes are then used to pump two $12\,cm$ PCF used as heralded single photon sources. This length of fibre was chosen so as to minimize the walk-off (which could lead to temporal distinguishability) between the pump and signal pulses. We chose to perform the interference experiment using the signal photons of each PCF. In order to get the spectral indistiguishablility (time-bandwith limited), a bandpass filter of width $\sim0.2\, nm$ (corresponding to the pump pulse duration) is used in the signal arm while a $\sim2\,nm$ width filter is used in the idler. As the photon pairs are matched in energy a narrowband filter on one photon of the pair will act as a non-local filtering on the other photon. The idler photons of each source are then launched into single mode fibres and the signal photons into a single-mode 50:50 coupler with a polarization controller (PC) set in one of its arms. All the outputs are finally connected to four Silicon avalanche photodiodes (APD) linked to a four-fold coincidence measurement apparatus (Ortec Quad 4-Input Logic Unit CO4020 + PC based counter card). The strong filtering process reduces significantly the lumped efficiencies. Nevertheless, we can use the brigthness of the source and increase the pump power to $P_{p}=8\,mW$ which compensate the loss and allows us to achieve a coincidence rate of $10^5\,/s$ within this narrow bandwith. 

\begin{figure}[h]
\centerline{\scalebox{0.9}{\includegraphics{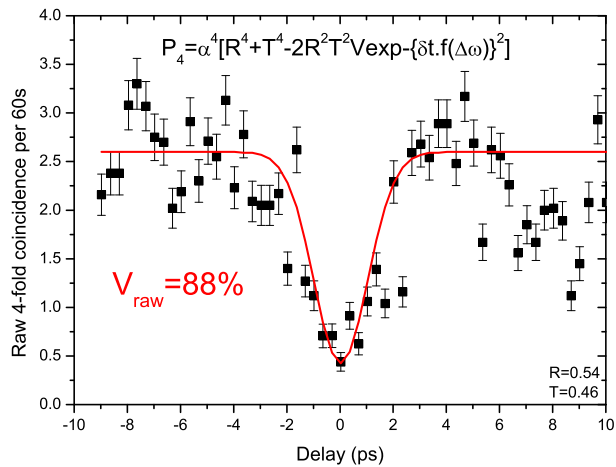}}}
\caption{Raw four-fold coincidence rate per 60s as a function of temporal overlap in ps.\label {dip1}}
\end{figure}

A retro-reflector mounted on a micrometric translation stage in one arm of the interferometer allowed us to scan through the temporal overlap $\delta T$ between the two signal photons. \fref{dip1} shows the raw four-fold coincidence rate as a function of the position of the retro-reflector for a very low pump power ($P_{p}=1.4\,mW$ per fibre). At $\delta T=0$ a significant reduction of the raw four-fold coincidence count rate was demonstrated. The term "raw" here means that no correction was applied on the result presented. The visibility of this dip was calculated taking into account the fact that our coupler is not perfectly 50:50. Using \eref{prob} (with $t=0.54$ and $r=0.46$) and fitting the experimental data allowed us to get back to the actual visibility of $88\%$. 

\begin{figure}[h]
\centerline{\scalebox{0.9}{\includegraphics{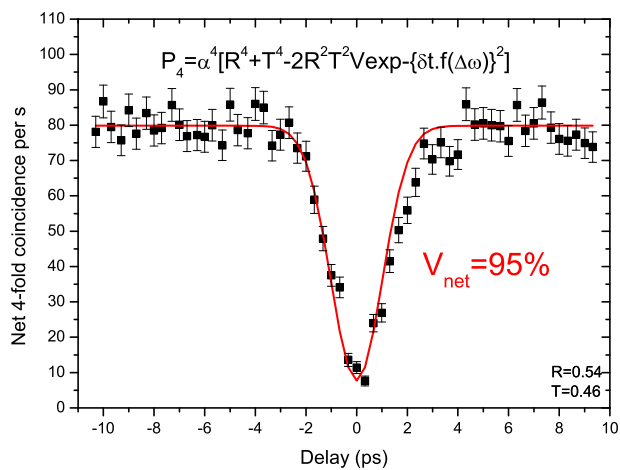}}}
\caption{Net four-fold coincidence rate as a function of temporal overlap in ps.\label {dip2}}
\end{figure}

\begin{figure}[h]
\centerline{\scalebox{0.9}{\includegraphics{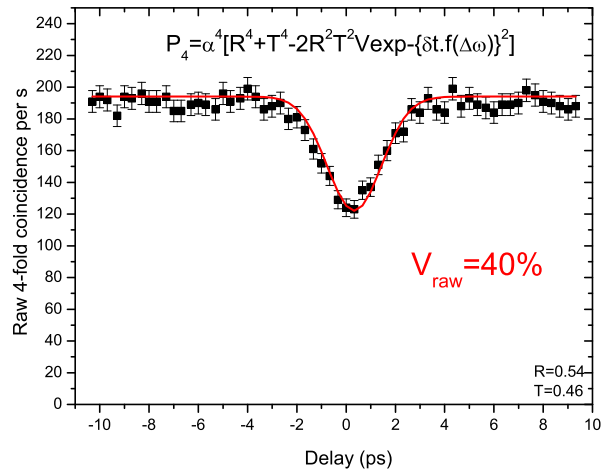}}}
\caption{Raw four-fold coincidence rate as a function of temporal overlap in ps. Taken from the same data set as \fref{dip2} but without background correction.\label {mandelraw40}}
\end{figure}

\begin{figure}[h]
\centerline{\scalebox{0.9}{\includegraphics{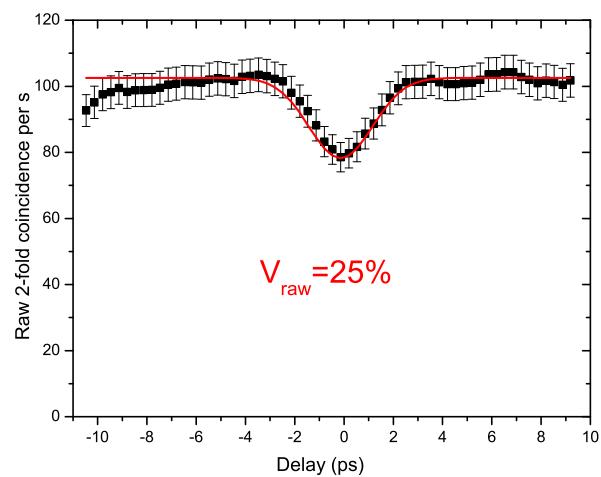}}}
\caption{Raw two-fold coincidence rate as a function of temporal overlap in ps. Result taken from the same data set as \fref{dip1} but without heralding.\label {mandel2fold}}
\end{figure}

We then increased the pump power to $P_{p}=8\,mW$ per fibre and plotted on \fref{dip2} the net four-fold coincidence rate as a function of the temporal overlap of the photons on the coupler. Unlike the previous results, this plot shows the net count rate which implies that the accidental four-fold coincidences ($\sim80$ per second per fibre) have been subtracted. Due to the high brightness of our source, which demonstrates 240 raw four-fold coincidences per second for $\sim8\, mW$ of pump power per fibre, multi-photon events lead to accidental four-fold coincidences. These events can be subtracted from the raw rate in order to get the true visibility of the interference. Experimentally this background is measured by blocking one input port of the coupler (and subsequently the other one) and looking at the four-fold coincidence rate coming from each fibre at a time revealing the amount of multi-photon events. Once corrected and considering again the defects of our 50:50 coupler, we could retrieve the actual visibility of the dip which came out to be $95\%$. 
This is close to the maximum theoretical visibility of $97\%$ we can get predicted from the filter widths (see \eref{vis}). Finally, beside this great visibility it is also worth noting the high net four-fold counting rate we can achieve of about $80\,/s$. This is comparable to the best coincidence rates achieved in bulk crystal multiphoton experiments \cite{6photpan} but could actually be greater since we are not limited in pump power. Although this result definitely highlights the quality of this source for quantum information purposes, we realise here that the post-selection method to make a single photon source is spoilt at high pump power. In \fref{mandelraw40} we show data of \fref{dip2} but without background correction. The raw visibility of $40\%$ is above the classical limit of $33\%$ for sources with Gaussian field statistics, which is induced by our filtering \cite{riedmatten:022301}. By ignoring the heralding entirely we get the result shown in \fref{mandel2fold} where we see $25\%$ visibility, below the maximum expected $33\%$.


\section{Discussion \label{disc}}
We can definitely achieve high raw visibilities at low pump power but at higher pump powers we are dogged by a background that arises mainly from cases where two pairs are created in one fibre and one pair in the other. In our data analysis we are able to measure this background and subtract it simply by alternately blocking the inputs to the beamsplitter. In the appendix we present a brief analysis of this background term and reserve a full analysis to a future paper.
In the limit of small $\overline n$ and low lumped efficiencies ($\eta_{s,\,i}$) then we have 
\begin{equation}
V\approx\frac{1+8 \overline n }{1+12\overline n }
\end{equation}

We then look at the rate of (non-interfering) fourfold coincidence detections and in the limit of small $\overline n$ then we have
\begin{equation}
C_4=R\overline n^2\eta_s^2\eta_i^2/2.  
\end{equation}
where $R$ is the laser repetition rate. \\

Clearly to get high raw visibility ($V>0.9$) we must limit $\overline n < 0.025$. At the same time we would like to use as high as possible $\overline n$ to obtain a high fourfold coincidence rate. We can increase four photon rate without reducing visibility by increasing the lumped efficiency of detection (in fact this also has a small effect on improving visibility, see appendix). We are also able to increase $C_4$ by increasing the repetition rate $R$ of our laser as we are not limited in output power. In our experiment effective detection efficiencies are $\eta_s\sim0.034$ and $\eta_i\sim0.05$ with laser repetition rate $R=8.2.10^7\,/s$ and limiting $\overline n\sim0.025$ we get $V\sim0.89$ and see a fourfold coincidence rate $C_4\sim \, 4/60s $ close to what we measure. 

In the future we can expect to increase efficiency to $0.1$ in both channels by use of improved filters, better coupling and by increased detector efficiency (using wavelengths closer to detector efficiency maximum) which would make 6-photon experiments more realistic. With an average number of pairs per pulse of $0.1$ and a repetition rate of $164\, Mhz$ (twice our current repetition rate) a raw six-fold coincidence rate up to $0.164\,/s$ would be obtained (before any beam-splitters). This coincidence rate would already be similar to the best rate reported so far in a 6-photon experiment \cite{6photpan}. Finally, in a more optimistic point of view where we could get efficiencies of $0.2$ in each channel, the previous result would be multiplied by $2^6$ leading to more than $10$ six-fold coincidences per second.

\section{Conclusion \label{conclu}}
In conclusion we have demonstrated a bright source of pure state photon pairs which can be used in multiphoton interference experiments showing high visibility interference between separate photons. We can easily generate hundreds of fourfold coincidence events per second and are effectively no longer limited by pump laser power. Our experiments are however limited by a strong background coincidence rate arising from pairs of pairs generated in each fibre. These higher order terms become significant when the average number of pairs generated per pulse reaches $\overline n\sim 0.025$ and are worsened by the Gaussian statistics of the narrowband filtered source. The short term solution to this problem is to increase the pulse repetition rate and reduce $\overline n$. Further improvement can be achieved by increasing the lumped collection and detection efficiency possibly by using better quality filters or developing an all fibre experiment. We can improve the heralding by using photon number resolving detectors to suppress the multi-photon events. We can also look at this phenomenon as a good indication that future 6-fold and even 8-fold photon coincidence experiments are possible. 

\appendix
\section{Background analysis}
We can write the state generated in the four wave mixing process up to two generated pairs  
\begin{equation}\label {state}
|\Psi\rangle=  \mathcal{N}[|0,0\rangle_{s,i}+\alpha|1,1\rangle_{s,i}+C{\alpha^2}|2,2\rangle_{s,i} + O(\alpha^3)]
\end{equation}
where again $|\alpha^2|=\overline n$ is the mean number of pairs generated per pulse and $\mathcal{N}$ is a normalising factor. The assumption that the generation process is random and Poisson would lead to constant $C=1/\sqrt{2!}$. However when we filter the state to make it indistinguishable we reach Gaussian statistics and $C=1$ \cite{RT98}. The idler photon is detected in a detector to herald the arrival of the signal photon. Our generic detector has lumped efficiency $\eta$ (including filter and transmission loss) and fires in response to an $n$ photon input with probability   
\begin{equation}\label{detector}
    P_{n}(1) = (1-(1-\eta)^n)
\end{equation}
After detection of one idler photon we can rewrite the heralded state as 
\begin{equation}\label {heraldedstate}
|\Psi\rangle_{s}=  \mathcal{N}'[|1\rangle_{s}+\sqrt{\frac {2\overline n\gamma}{1+2\overline n\gamma}}|2\rangle_s + O(\alpha^2)]
\end{equation}
where $\gamma=(1-\eta_i/2)$ and normalisation is by $\mathcal{N}'$. At this stage we assume no losses in the signal arm and apply all losses later as lumped detector efficiencies.

We can then generalise the beamsplitter transformation to arbitrary number states at both inputs 
\begin{equation}\label{nmstatebs}
 |n\rangle_1 |m\rangle_2 \rightarrow
 (t a_3^\dagger + ir a_4^\dagger)^n  (ir a_3^\dagger + t a_4^\dagger)^m
 |0\rangle_3|0\rangle_4 / \sqrt{n!m!}
\end{equation}

  $=\sum_{j=0}^n\sum_{k=0}^m
   t^{m-k+j} (ir)^{n-k+j} \sqrt{C_j^nC_k^mC_j^{(j+k)}
  C_{(n-j)}^{(n+m-j-k)}} |(j+k)\rangle_3 |(n+m-j-k)\rangle_4$\\
where $a_k^\dagger$ is the creation operator in the mode $a_k$ with $k= 3,$ $4$.

We can apply now the two inputs of heralded single photons as expressed by \eref{heraldedstate} using \eref{nmstatebs} and then calculate the probability of seeing a coincidence detection in the outputs $a_3,a_4$. In the ideal case this would be zero but here we see
\begin{equation}\label{interf}
P(\ket{1}_3,\ket{1}_4)_{int}=\eta_s^2[\frac {2\overline n\gamma\gamma'}{1+2\overline n\gamma}]
\end{equation}
to first order in $\overline n$ and with $\gamma'=(1-\eta_s/2)$.   
When the photons do not overlap we must apply separate beamsplitters (\eref{nmstatebs}) to each heralded state with $n/m=0$ in the second port and then sum the probabilities in the outputs to obtain
\begin{equation}\label{noninterf}
P(\ket{1}_3,\ket{1}_4)_{noint}=\eta_s^2[\frac{1}{2}+\frac {6\overline n\gamma\gamma'}{1+2\overline n\gamma}]
\end{equation}
As we define the visibility as 
$V=\left[P(\ket{1}_3,\ket{1}_4)_{noint}-P(\ket{1}_3,\ket{1}_4)_{int}\right]/P(\ket{1}_3,\ket{1}_4)_{noint}$ then we find 

\begin{equation}
V\approx\frac{1+8 \overline n \gamma\gamma'}{1+12\overline n \gamma\gamma'}
\end{equation}

\end{document}